%% file: main.tex
\documentclass[conference]{IEEEtran}
\usepackage{cite}
\usepackage{url}
\usepackage[inline]{enumitem}
\usepackage{grffile}
\usepackage[ruled,vlined]{algorithm2e}

\usepackage{amsmath,amssymb,amsfonts}

\usepackage[english]{babel}
  
\usepackage{nomencl}

\SetCommentSty{mycommfont}
\usepackage{graphicx}

\usepackage{textcomp}
\usepackage{xcolor}
\usepackage{lipsum}
\usepackage{xfrac}
\usepackage{commath}
\usepackage{flexisym}
\usepackage{makecell}
\usepackage{mathtools}
\usepackage{multicol}
\usepackage[labelfont=bf]{caption} 
\usepackage{caption}
\usepackage[section]{placeins}
\usepackage{cleveref}

\def\BibTeX{{\rm B\kern-.05em{\sc i\kern-.025em b}\kern-.08em
T\kern-.1667em\lower.7ex\hbox{E}\kern-.125emX}}

\SetKwComment{Comment}{$\triangleright$\ }{}
\makeatletter
\def\ps@IEEEtitlepagestyle{%
  \def\@oddfoot{\mycopyrightnotice}%
  \def\@evenfoot{}%
}
\def\mycopyrightnotice{%
  {\hfill \footnotesize 978-1-6654-4331-9/21/\$31.00 \copyright 2021 IEEE\hfill}
}
\makeatother

\begin{document}

\title{BigBFT: A Multileader Byzantine Fault Tolerance Protocol for High Throughput}

\author{\IEEEauthorblockN{Salem Alqahtani}
\IEEEauthorblockA{\textit{Computer Science Department} \\
\textit{University at Buffalo,SUNY}\\
salemmoh@buffalo.edu}
\and
\IEEEauthorblockN{Murat Demirbas}
\IEEEauthorblockA{\textit{Computer Science Department} \\
\textit{University at Buffalo,SUNY}\\
demirbas@buffalo.edu}
}
\maketitle

\input{Abstract}
\input{Introduction}
\input{Background}
\input{Preliminaries}
\input{Protocol}
\input{Correctness}
\input{Evaluation}

\input{Discussion}
\input{Futurework}
\input{Conclusion}
\input{Akc}

\bibliographystyle{IEEEtran}
\bibliography{Survey.bib}
\end{document}

%% file: Abstract.tex
\begin{abstract}

This paper describes BigBFT, a multi-leader Byzantine fault tolerance protocol that achieves high throughput and scalable consensus in blockchain systems. BigBFT achieves this by (1) enabling every node to be a leader that can propose and order the blocks in parallel, (2) piggybacking votes within rounds, (3) pipelining blocks across rounds, and (4) using only two communication steps to order blocks in the common case.
  
BigBFT has an amortized communication cost of $O(n)$ over $n$ requests. We evaluate BigBFT's performance both analytically, using back-of-the-envelope load formulas to construct a cost analysis, and also empirically by implementing it in our PaxiBFT framework. Our evaluation compares BigBFT with PBFT, Tendermint, Streamlet, and Hotstuff under various workloads using deployments of 4 to 20 nodes. Our results show that BigBFT outperforms PBFT, Tendermint, Streamlet, and Hotstuff protocols either in terms of latency (by up to $70\%$) or in terms of throughput (by up to $190\%$). 
  
\textbf{Keywords}: Byzantine fault tolerance, permissioned blockchain, distributed ledgers consensus.
\end{abstract}

%% file: Introduction.tex
\section{Introduction}

The proliferation of cryptocurrency boosted the adoption of Byzantine fault-tolerance (BFT) in many permissioned blockchain systems. Compared to Proof-Of-Work approaches~\cite{Bitcoin}, BFT protocols provide the advantages of computation efficiency and near-instant block finality.

BFT has been well-studied in the context of distributed systems~\cite{PBFT}. Tendermint~\cite{Tendermint} arose later and implemented PBFT like protocol for permisssioned blockchain systems. Unfortunately, both PBFT and Tendermint are communication heavy protocols and only applicable to small-scale systems. When extending the these protocols to a large-scale, its performance may become unacceptable.

Many BFT consensus protocols emerged recently to improve communication efficiency. These include Hotstuff~\cite{Hotstuff}, Streamlet~\cite{Streamlet}, Fast-Hotstuff~\cite{FHS}, and SBFT~\cite{golan2018sbft}. Unfortunately, these protocols still have limited scalability and performance because all communication go through a single leader, which constitutes some of the throughput bottlenecks~\cite{alqahtani2021bottlenecks,gai2021dissecting,ailijiang2019dissecting,yang2018linbft}. The leader performs a disproportionately large amount of work compared to its $N-1$ followers in some BFT protocols. This affects efficiency, scalability, and prevents high throughput when $N$ is increased. For example, in Hotstuff, for each consensus instance, the followers receive one message from the leader and send one message back. For instance, in a cluster of $N=4$ nodes, the leader handles $16$ messages for one consensus instance while followers handle $8$ messages in Hotstuff. To scale the system and to increase the throughput, we need to reduce the disproportionate workload on the leader and number of messages sent/received for one consensus instance.

To alleviate the single leader bottleneck, multiple leader protocols were introduced~\cite{Mir,avarikioti2020fnf,gupta2021rcc,barcelona2008mencius}. Mir-BFT is the first protocol that enables leaders to operate independently on different sequence spaces and reach consensus as long as there are no conflicts. However, the protocol inherits PBFT's message complexity and does not use pipelining techniques. Mir-BFT performance is reduced due to view change protocol.

In this paper, we design BigBFT, a parallel-leader BFT protocol that addresses these bottlenecks and shortcomings. BigBFT eliminates the single leader bottleneck and distributes the load over all leaders. Similar to Mir-BFT~\cite{Mir}, BigBFT rotates the coordinator to assign sequence space to each leader in the coordination phase.
Unlike Mir-BFT~\cite{Mir}, BigBFT pipelines the coordination phase to detect Byzantine coordinators by using a separate consensus/configuration box. This coordination phase is not on the critical path of the BigBFT protocol and are done concurrently with the previous round of the BigBFT consensus.
For improving the protocol scalability, the coordination phase signs a batch of sequence number to each leader and all leaders monitor the protocol in every sequence number.
After a new coordinator proposes a new batch of sequence numbers in the coordination phase, leaders agree on all blocks in parallel in BigBFT protocol. This way, BigBFT leaders can enforce the total order on all requests. A malicious leader in BigBFT impacts only itself and due to an incentive mechanism, the malicious leader has no incentives to do so. 
BigBFT has two communication steps to order blocks. Leaders propose the client requests for every other leaders. Upon receiving the $N-1$ propose messages, the leader piggybacks them together into a single voting message. Upon receiving a quorum certificate of the $N-F$ nodes, the node commits the value and replies to the client in the second round.
We model check the correctness of the high-level protocol in TLA+.

BigBFT pipelines commit of previous round with proposing requests of next round and amortizes the cost of voting phase over number of requests to reduce message complexity to be linear in BigBFT. We provide a back-of-the-envelope performance analysis of BigBFT, and provide experimental results and evaluate BigBFT in both LAN and WAN deployments.
BigBFT has an amortized communication cost of $O(n)$ over $n$ requests. We evaluate BigBFT's performance both analytically, using back-of-the-envelope load formulas to construct a cost analysis, and also empirically by implementing it in our PaxiBFT framework. Our evaluation compares BigBFT with PBFT, Tendermint, Streamlet, and Hotstuff under various workloads using deployments of 4 to 20 nodes. Our results show that BigBFT's latency is $70\%$ better than HotStuff, $55\%$ better than PBFT, and $65\%$ better than Streamlet. BigBFT in WAN can provide $100\%$ higher throughput than PBFT, and $190\%$ higher throughput than Streamlet, and can match HotStuff's throughput.

We implement BigBFT and compare and benchmark PBFT, Tendermint, Streamlet, and Hotstuff. Our experiments, conducted on AWS EC2 nodes with 4 to 20 nodes in various LAN and WAN topologies, show that BigBFT is effective in scaling consensus to large clusters. For 20 nodes LAN deployment, PBFT throughput gets saturated at 500 requests per second, Tendermint throughput reaches its limit of around 130 req/sec, Streamlet throughput reaches its limit of around 700 req/sec, whereas BigBFT scales beyond 2150 req/sec with little latency increased. HotStuff throughput reaches around 2400 req/sec, but with $40\%$ higher latency than BigBFT. 

%% file: Background.tex
\section{Background and Related Work}
\label{sec:BG}
\subsection{State Machine Replication}

State Machine Replication (SMR)~\cite{schneider1990implementing,lamport2019byzantine,pease1980reaching} is a general method for building fault tolerant systems. In SMR, every node stores a state of the system and applies the same set of commands on the state even if a fraction of them are faulty. In practice, SMR uses consensus protocols~\cite{pease1980reaching,lamport2019byzantine,PBFT,Hotstuff,Tendermint} to reach consensus among all nodes on the system state.

Consensus protocols guarantee $Non-triviality$ (the decided value $v$ was proposed by a correct node), $Safety$ (all correct nodes output the same value $v$), and $Liveness$ (eventually all correct nodes output some value). 

The famous FLP impossibility result~\cite{FLP} proved that a deterministic agreement protocol in an asynchronous systems cannot guarantee liveness if one node may crash. Many consensus protocols have been proposed to circumvent the FLP impossibility to achieve an asynchronous consensus such as failure detectors, randomness, and time assumptions. One example of consensus protocol that deals with FLP impossibility result to tolerate Byzantine nodes is PBFT~\cite{PBFT}(details can refer to Section~\ref{sec:pbft}). PBFT guarantees safety and liveness in the partial synchronous network model~\cite{dwork1988consensus}, and this is achieved under the $1/3$ optimal resilience bound~\cite{ben1983another}. 

\subsection{Leader Bottleneck}

Consensus protocols rely on a strong single leader to coordinate and to order the client requests in the system. This strong leader, however, is often a bottleneck, especially when every read and write operation has to go through it. A strong leader needs to send messages to all the replicas, and receive responses to know when the operation has been successfully replicated in the state machine. In our recent work in BFT protocols, we identified and studied single-leader bottlenecks~\cite{alqahtani2021bottlenecks}. In the more common case, the leader will be bottlenecked at the CPU serializing, deserializing, and processing these messages. Too many messages are sent, received, and processed by one node. To solve leader bottlenecks in a total ordering, we alleviate the single leader bottlenecks by using multi-leaders. 

\subsection{Single Leader}
\subsubsection{PBFT}
\label{sec:pbft}
PBFT protocol~\cite{PBFT} provides the first practical solution to the Byzantine problem~\cite{lamport2019byzantine}. PBFT employs an optimal bound of $\!N\!\!\!\geq\!\!3F\!\!+\!\!1\!$ leaders, where the Byzantine adversaries can only control up to $\!F\!$ leaders. PBFT uses encrypted messages to prevent spoofing and to replay attacks, as well as to detect corrupted messages. PBFT employs a leader-based paradigm, guarantees safety in an asynchronous model, and guarantees liveness in a partially synchronous model. PBFT requires $O(n^2)$ transmissions in its best case and $O(n^4)$ in the worst case.

\subsubsection{Tendermint BFT}
Tendermint~\cite{Tendermint}, used by Cosmos network~\cite{cosmos}, utilizes a proof-of-stake for leader election and votes on appending a new block to the chain. Tendermint rotates its leaders using a predefined leader selection function that priorities selecting a new leader based on its stake value. The protocol employs a locking mechanism after the first phase to prevent any malicious attempt to make leaders commit different transactions at the same height of the chain. Each leader starts a new height by waiting for prepare and commit votes from $2F+1$ leaders and relies on the gossip network to spread votes among all leaders in both phases. Tendermint prevents the hidden lock problem~\cite{Tendermint} by waiting for $\delta$ time. The hidden lock problem occurs because receiving $\!N\!-\!F\!$ replies from participants (up to $\!F\!$ of which may be Byzantine) alone is not sufficient to ensure that the leader gets to see the highest lock; the highest lock value may be hidden in the other $\!F\!$ honest nodes which the leader did not wait to hear from. Such an impatient leader may propose a lower lock value than what is accepted and this in turn may lead to a liveness violation.

\subsubsection{HotStuff BFT}
\label{sec:hs}
HotStuff protocol~\cite{Hotstuff}, is used in Facebook's Libra~\cite{Librag}. HotStuff rotates leaders for each block using a rotation function. HotStuff is responsive; it operates at network speed by moving to the next phase after the leader receives $N-F$ votes. This is achieved by adding a pre-commit phase to the lock-precursor. To assign data and show proof of message reception and progression, the protocol uses Quorum Certificate(QC), which is a collection of $N-F$ signatures over a leader proposal. Moreover, HotStuff uses a one-to-all communication. This reduces the number of message types and communication cost to become linear. The good news is that, since all phases become the same communication-pattern, HotStuff uses a pipeline mechanism and performs four leader blocks in parallel, thus improving the throughput by four.

\subsubsection{Streamlet}
\label{sec:Streamlet}

Streamlet protocol is a consensus algorithm that was proposed in 2020~\cite{Streamlet} as a simpler alternative to PBFT-based blockchain protocols. Streamlet leverages the blockchain infrastructure and the longest chain rule in the Nakamoto protocol~\cite{Bitcoin} to simplify consensus. Similar to both Tendermint and HotStuff, Streamlet rotates its leader for each block using a rotation function. The protocol proceeds in consecutive and synchronized epochs where each epoch has a dedicated leader known by all validators. Each epoch has a leader-to-participants and participants-to-all communication pattern. This reduces the number of message types, but the communication cost is not linear $O(N^3)$. Streamlet has a single mode of execution and there is no separation between the normal and the recovery mode. Streamlet guarantees safety even under asynchronous environments with arbitrary network delays and provides liveness when synchrony assumptions start to hold.

\subsection{Multi-Leaders}

\subsubsection{Mir-BFT}
\label{sec:Mir-BFT}

Mir-BFT~\cite{Mir} is a multi-leader consensus protocol that aims to improve the scalability and throughput of the system. Mir-BFT starts by partitioning the request hash space among all leaders to solve duplication attacks and rotates request hash space among all leaders to solve censorship attacks. It also uses batching and watermarks to facilitate concurrent proposals of batches by multiple parallel leaders. Mir-BFT proceeds in epochs and each epoch has a single primary and a set of leaders. Each leader will run an independent instance of PBFT~\cite{PBFT}. Mir-BFT improves the performance throughput in WAN deployment and introduces a more robust BFT protocol. In terms of leaders failure, the throughput can only recover after multiple view changes discard the faulty leaders. If many leaders suspect the primary, then they timeout the epoch, and ask for epoch to be changed.
\subsubsection{FnF-BFT}
\label{sec:FNF}
FnF-BFT~\cite{avarikioti2020fnf} is a parallel-leader BFT consensus protocol that provides high throughput under malicious behaviors. FnF-BFT uses a Byzantine resilient performance metric to evaluate a BFT's performance. FNF-BFT has view change protocol, but has linear communication complexity during the synchrony. FnF-BFT can achieve Byzantine-resilient performance with a ratio of $16/27$ while maintaining both safety and liveness. FnF-BFT provides three properties under a stable network which are optimistic performance, Byzantine-resilient performance, and efficiency. To achieve these three properties, FnF-BFT enables all replicas to continuously act as leaders in parallel to share the load of client’s requests and does not replace leaders upon failure but based on the performance history.

\subsubsection{RCC}
\label{sec:RCC}
RCC~\cite{gupta2021rcc} is concurrent leader protocol that introduced a paradigm called RCC that enables any message exchange patterns to run in parallel. The protocol requires instances to unify after each request creating a significant overhead. Additionally, the protocol relies on failure detection, which is only possible in synchronous networks. With BigBFT, we allow leaders to make progress independently of each other without any affect of failure detection.

%% file: Preliminaries.tex
\section{Preliminaries}
\label{sec:Preliminaries}

\subsection{System Model}

We consider a permissioned blockchain system with arbitrary number of clients and a finite set of leader nodes $\!N\!=3F\!+\!1$, indexed by $i\in \{1,...,n\}$, where leaders can tolerate up to $F$ Byzantine leaders. Leaders maybe in multiple geographical locations and have both different speeds and different physical machines. All leaders communicate and synchronize by sending and receiving messages through reliable channels. A correct leader follows its specification while Byzantine leaders control by an adversary and behave arbitrary including sending wrong messages and colluding with each other to harm the system. A computationally bounded adversary can control the faulty nodes to compromise the system if more than $F$ compromised. 

BigBFT protocol runs in rounds and each round consists of two phases called propose and vote phases. We assume that at the beginning of a round, every leader knows all other leaders in the round. The coordination phase is not on the critical path of the protocol. In each round, we assumes a unique coordinator is known to every leader for every round. The coordinator node can be one of the leaders. We also assume a round-robin rotation for coordinator elections as in many BFT protocols~\cite{Streamlet,Hotstuff,Tendermint,gupta2021rcc,avarikioti2020fnf,Mir}.

PBFT has a complicated view-change sub-protocol with a quadratic message complexity. Chain based protocols~\cite{Tendermint,buchman2016tendermint,buterin2017casper,Hotstuff,Streamlet,shi2019streamlined} have emerged recently that simplify view-change sub-protocol to have linear message complexity. However, chain based protocols requires sequential order to propose and commit proposals which affect concurrency executions, which lead to low resource utilizations. Fortunately, BigBFT coordination phase is not on the critical path of the protocol and both BigBFT and its coordination phase are working in parallel to maximize resource utilizations. 

\subsection{Communication Model}

Following common practice in the literature, we assume a partial synchrony communication model in our protocol~\cite{dwork1988consensus}, as most BFT protocols of the same kind~\cite{PBFT,Streamlet,chan2018pala,golan2018sbft,Hotstuff}, where there is a known network delay bound $\delta$ that will hold after an unknown Global Stabilization Time (GST). After GST, all messages between honest leaders will arrive within time $\delta$. When an honest leader sends a message in round $r$, an honest recipient leader is guaranteed to receive it by the beginning of round $(r + \delta)$. Although we assume partial synchrony, the protocol achieves consistency (i.e., safety) regardless of how long the message delays are or how badly the network might be partitioned. The protocol executes in parallel with a linear communication complexity over $N$ nodes. Also, the protocol maintains responsiveness which proceeds as network delivers~\cite{Hotstuff,hu2021don}.  

\subsection{Cryptographic Primitives}
\label{sec:DS}

In this part, we introduce the cryptographic algorithms used in BigBFT. We take the advantages of existing cryptographic tools that are available. Similar to many BFT protocols~\cite{Hotstuff,abraham2018hot,golan2018sbft,yang2018linbft}, we assume standard digital signatures and public-key infrastructure (PKI) that identify all leader and client processes. The BigBFT's message exchange patterns combined with some cryptographic primitives to create digital signatures. BigBFT uses a signature aggregation scheme~\cite{boneh2003aggregate} that reduces the message complexity~\cite{berger2018scaling} and enables leaders to convert a set of signatures into a single signature; however, this can only happen when the set contains threshold value which is $N-F$ in BigBFT.

The scheme allows leaders to receive $N-F$ partial signatures $\sigma_{i}=sign_{i}(B_{j})$ from every leader $i$ for every block $B_{j}$, and combine them into a single signature  $\sigma =$AggSign(sign$_{i}(B_{j})$ $_{i \in N}$ and $_{j \in B_{j}}$). The leader then aggregate all $\sigma$ in a single signature $AggQC$. 

BigBFT uses message digest to detect corrupted messages. Both clients and leaders must be able to verify each other's leader public key and messages. We assume that all cryptographic techniques cannot be broken. We also assume cryptographic hash function $H(.)$ that maps arbitrary input to a fixed size output. We assume the hash is collision resistant where is no $H(x) == H(y)$. 

%% file: Protocol.tex
\section{BigBFT Protocol}
\label{sec:Protocol}

\subsection{Overview}

BigBFT protocol has a designated coordinator $C_{r}$ that has chosen in a round-robin fashion for leading the round $r$, managing the leader set $L_{s}$, and partitioning the set of sequence numbers $\mathbb{Z}$ between leaders to avoid conflicts between leaders, $Partition_{{k}{i}} \gets \text{$\mathbb{Z}/L_{s}$}$. The $Partition_{{k \in \mathbb{Z}}\text{ and }{i \in L_{s}}}$ means a partition $k$ assigned to a leader $i$ from the leader set $L_{s}$. 

\begin{figure}
	\centering
	\includegraphics[width=3.5in]{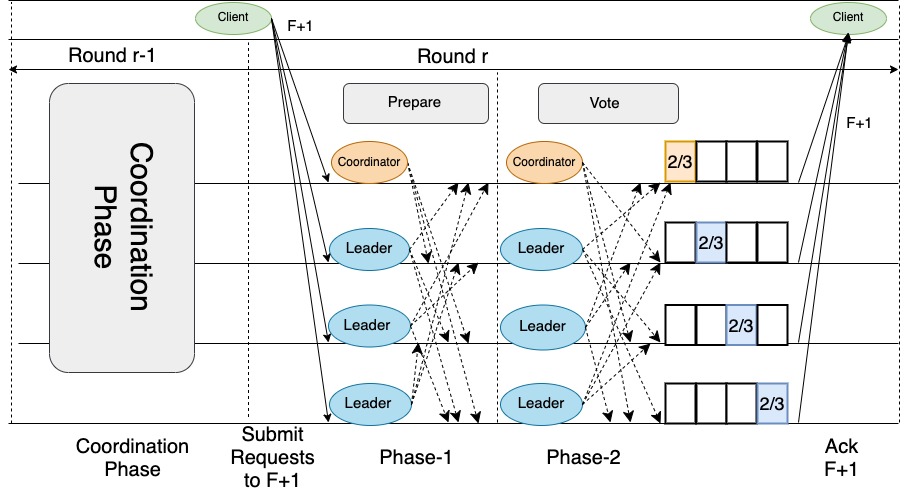}
	\caption{\textbf{Communication pattern of BigBFT}}
	\label{fig:BigBFT}
\end{figure}

BigBFT executes in rounds $r= [0,+\infty)$ where each round has a dedicated leader set and a coordinator known to all. The communication patterns when the network is synchronous, no byzantine failures, and no contention, are two communication phases in a normal-case: \begin{enumerate*}[label=(\roman*)] \item leaders propose client requests in parallel(Phase-1). \item vote on client requests in parallel(Phase-2)\end{enumerate*}. In Figure~\ref{fig:BigBFT}, we illustrate the communication flow of BigBFT protocol. 
The BigBFT would only perform coordination phase(round-change) after many consensus instances in parallel with BigBFT protocol(Phase-1 and Phase-2). Coordination phase invokes to replace a Byzantine coordinator, a leader set, and a new partition of the sequence numbers $\mathbb{Z}$.

Any full node in the system have two roles called coordinator ($C$) and leader ($L$). $C$ is responsible for leading the round and partitioning the sequence number space. $L$ is responsible to receive a partition space, proposing blocks and voting on blocks. These roles can be co-located, that is, a single process can be a coordinator or a leader. 

Clients submit their requests to $F+1$ leaders in the system that are responsible to handle the client request. In this case, non-faulty process can learn the request if the faulty leader tries to prevent that request from being in the next proposal. BigBFT's $client$ can send many independent requests on the fly to the leaders $\big \langle \! Request, t, O, id\big \rangle$, where $t$ is the timestamp, $O$ is the operation, $id$ is the client id. The client will receive $\big \langle \! Reply, r, t, L\big \rangle$, where $r$ is the round number and $L$ is the leaders identifications who executed the client request.

  
        
  
\begin{figure}
	\centering
	\vspace*{-4mm}
	\includegraphics[width=3.5in]{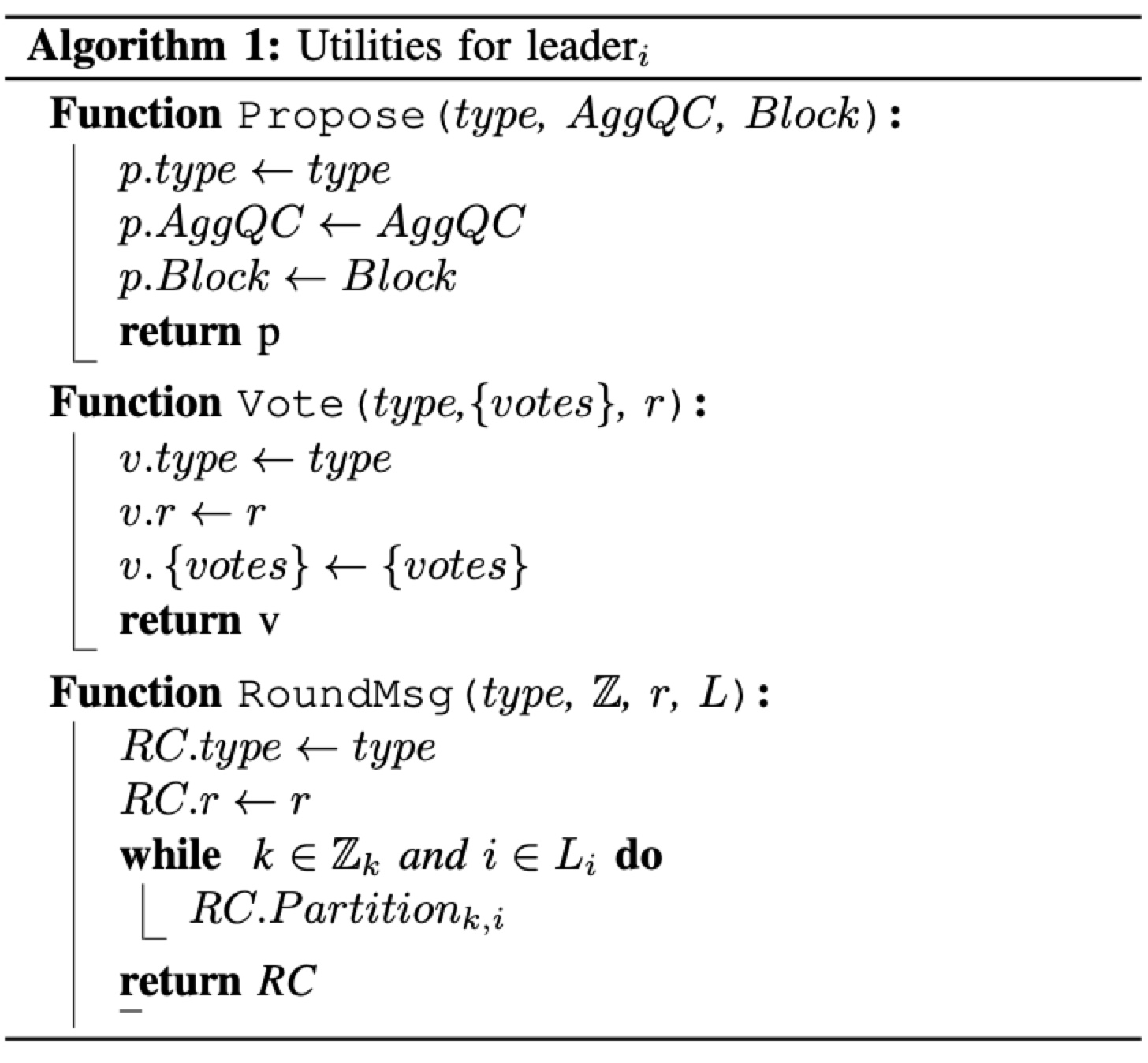}
	\vspace*{-7mm}
	\label{alg:alg1}
\end{figure}

\subsection{\textbf{Data Structures}}

This part introduces message types and the structure of the block in BigBFT before presenting the details of the protocol. As shown in the Algorithm~\ref{alg:alg2}, $RChange$ is the round change message to change the current round that carries a new round number $r = r+1$ where $r$ is the current round, space partitions $\mathbb{Z}$, and a leader set $L_{s}$. 

$Prepare$ message in the Algorithm~\ref{alg:alg3} carries the digest message of block $B_j$, the sequence number $sn$, the round $r$, and the previous round $AggQC_{r-1}$. The leaders always use $AggQC_{r-1}$ to proof the last blocks commit to propose a new block.
$Vote$ message in the Algorithm~\ref{alg:alg3} carries a set of partial signatures of blocks for $B_{j}$ where $j \in B_{j}$. An aggregate quorum consists of set of $QCs$ that each $QC$ contains $N-F$ signed votes for a block from distinct leaders. 

\begin{algorithm}
  \caption{Coordination phase of BigBFT}
  \label{alg:alg2}
  \DontPrintSemicolon
  \Comment*[l]{Coordination phase}
  \ForEach { $r \gets 0,1,2... $ }{

    \If{ $C_i$ is coordinator}{
	RC $ \gets RoundMsg(RChange,\mathbb{Z},r,L)$\;
    	\text{Broadcast\text{ RC }}\;
	}
    
    \If{ $L_i$ is leader}{
    \If{RC is received from $C_i$ }{
    	\If{( $r >=$ local r )}{
	 $\sigma_{i}=Sign(Ack,sk_i)$\;
	         \tcp{sk = secret key of leader i}
    		Send $\sigma_{i}$ to $C_i$}
    }}
     \If{ $C_i$ received (N-F $\sigma_{i}$ where $i \in N$ ) in r}{
     $RoundQC \gets$ \text{combine N-F $\sigma_{i}$ where $i \in N$}\;
     broadcast\text{ RoundQC to N-1 }}
     
   \If{($L_i$ receives an RoundQC from  $C_i$)}{
       Call Algorithm~\ref{alg:alg3}}
  }
\end{algorithm}

\subsection{Coordination Phase}

We design the coordination phase as a separate phase of BigBFT to avoid any leader bottlenecks. The coordination phase partitions the sequence space across replicas in $r$ and prepares leader set $L_{s}$ for next round $r+1$. In each round, the new coordinator increases the round number by one and starts the round, $C_{i+1} \leftarrow \text{$next.C_{i}$}$. As described in Algorithm~\ref{alg:alg2}, the coordinator $C_{i+1}$ sends round-change $RChange$ message to all leaders. Each leader receives and processes the $RChange$ against its state. Then, each leader signs the coordinator message $RChange$ and sends the reply back to coordinator. Upon receiving $N-F$ replies from leaders, the coordinator creates $RoundQC$ quorum and broadcasts it to all leaders in round $r$. The coordination phase executes in parallel with BigBFT~\ref{sec:subag} of previous round $r-1$. After the execution of coordination phase, each leader will have a designated partition of sequence numbers assigned by the coordinator.

\begin{algorithm}
  \caption{BigBFT protocol}
  \label{alg:alg3}
\DontPrintSemicolon
 \For{ each r $\gets 0,1,2...$} {
  \Comment*[l]{Prepare phase of $L_i$}
  
    $l_i \Big \langle$ waits for $B_j \Big \rangle$ from client Then\\
     \If{$B_j == valid$}{
     $msg = Propose(\text{prepare}, AggQC_{r-1}, B_{j})$ \;
     $broadcast(msg, d(B_j),r) $\;
     }
   \Comment*[l]{$L_i$ receives prepare Msgs}

   \If{ receives N-F prepare msgs}{

   \tcp{Same block cannot be assign to more than one leader}

   \While{ $r=true$} {
     \If{$B_j.L_i == B_j.L_i+1$}{
        $r \gets \text{False}$}
        }
     \If{$AggQC_{r-1} == true$}{
        $((N-F)B_j \in r-1) \gets \text{committed}$}} 
        
   \Comment*[l]{Send Vote Phase of $L_i$}

     \If{ $L_i$ received (N-F $B_j$ where $j \in N$) in r}{
        $\set{\sigma} = \bigcup_{j=1}^{N-F} \sigma_{j}=Sign_i(B_{j})$\;
        broadcast Vote(vote,\text{\set{\sigma}}, r)
        }
   \Comment*[l]{$L_i$ receives Vote Msgs}
	$QC(B_{j}) \gets \bigcup_{i=1}^{N-F} \sigma_{i}$\;
	$AggQC_{r} \gets \bigcup_{i=1}^{N-F}$ \text{QC$_i$}\;
	}
\end{algorithm}

\subsection{BigBFT protocol}
\label{sec:subag}

After partitioning the set of sequence numbers across leaders in coordination phase, BigBFT starts two communication phases called prepare and vote. Below we describe how prepare and vote phases works. 

\textbf{Prepare Phase.} Upon receiving the block in round $r$, each leader assigns the next available sequence number to the new block. The leader also attaches the proof of vote $AggQC$ from previous round $r-1$ in prepare message. Then, in prepare phase, each leader proposes the prepare message to all leaders.

In normal path, the leaders receive $N-F$ proposed blocks that do not conflict with any other blocks. Every leader checks the $AggQC_{r-1}$ in order to commit the blocks from previous round $r-1$. If the leader is faulty, the BigBFT guarantees that the faulty leader affect only its process and pending blocks go to other honest leaders.

\textbf{Vote Phase.} Upon receiving $N-F$ proposed blocks, the leader signs each block and sends the vote message to all leaders. To reduce the message complexity in the vote phase, instead of sending vote messages $N-F$ times, we combine all vote messages in a single vote set $\set{votes}$ and send it to all leaders. Each vote message represents a partial signature that signed by replica $i$ for a block. Because we have $N-F$ blocks, we need to combined $N-F$ signatures $\set{\sigma}$ in one vote message.  

We describe the BigBFT protocol as follow.

\begin{enumerate}

\item Client nodes broadcast their blocks/requests to $F+1$ leaders including the coordinator. 
\item Upon receiving new block proposal from clients, every leader verifies the block, assigns sequence number to the new block, and attaches the proof of vote $AggQC$ from previous round in the prepare message. 
\item Leader broadcasts prepare message to all other leaders. 
\item Upon receiving prepare messages and commit $AggQC_{r-1}$, leaders commit previous round blocks. Leaders sign each proposal message and combine all signatures in a vote set $\set{votes}$. 
\item Upon receiving the $N-F$ vote set messages, leader create a quorum certificate for every block and combined all quorum certificate in aggregated quorum certificate $AggQC$, Algorithm~\ref{alg:alg3}. 
\item In the next round $r+1$, leader nodes check the new propose blocks. If the prepare message carries the $AggQC_{r}$, then it commit the blocks and reply to the client. This is what we have called across rounds pipelining.
\end{enumerate}
Importantly, the entire protocol follows a unified coordination-prepare-vote paradigm. The coordination protocol is pipelined and not on the critical path of BigBFT.

%% file: Correctness.tex
\section{BigBFT Correctness}
\label{sec:proof}

We prove that BigBFT achieves both safety and liveness properties by showing BigBFT algorithm solves agreement, validity, and termination in all possible distributed executions. We also performed model checking of the high level BigBFT protocol in $TLA^+$~\cite{lamport2002specifying}. The specification is available on the GitHub.\footnote{\url{https://github.com/salemmohammed/BigBFT/tree/main/tla}}

\subsection{Safety}

BigBFT guarantees its safety in any circumstances regardless of the network delays and partitions. If there are less than $\!F\!\!\!<\!\!\!\frac{N}{3}\!$ Byzantine leaders and $\!\!N-F\!\!$ honest leaders decide on blocks $B_{j \in(j,j+N-1)}$ at blockchain height $\{$h$ \text{ where $h=h$ to $h=h+N-1$}\}$, then no honest leader will decide on any blocks other than $B_{j}$. 


\textbf{Lemma 1} If we have $AggQC_{1}$ and $AggQC_{2}$ with $F\!\!<\!\!\frac{N}{3}$, then both $AggQCs$ are not on the same round $r$.

$Proof.$ Suppose that blocks from $B_j$ to $B_{j+N-1}$ are committed in $AggQC_{1}$ at round $r$ and blocks from $B_{j+N}$ to $B_{j+2N-1}$ are committed in $AggQC_{2}$ at round $r'$. The number of blocks are determined by the number of active leaders who propose blocks in parallel. It must be that at least $N-F$ leaders denoted as $S_{0}$, signed the block $B_{j \in(j,N-1)}$, and at least $N-F$ leaders denoted as $S_{1}$, signed the blocks $B_{j\in(j+N,j+2N-1)}$. Since there are only $N$ leaders in total, $S_{0}\cap S_{1}$ must intersect in at least $\frac{1}{3}$, and thus at least one honest leader is in $S_{0}\cap S_{1}$. According to our protocol, every honest leader votes for at most one time for each height in the blockchain. Therefore, it must be that $r \neq r'$ and $AggQC_{1}$ is a prefix of $AggQC_{2}$.

\textbf{Lemma 2} If at least one correct leader has received $N-F$ votes for block $B_j$ in round $r$, then if some leaders have increased their round numbers due to network partitions or node failures, the round $r$ is the last round that have the latest valid votes before proposing the next round.

$Proof.$ The leaders receive client's request in round $r$. The leader also have $N-F$ votes from each request in previous round $r-1$. When each leader proposing both the new requests and the $AggQC$ for $r-1$ in round $r$, the leaders receive proofs for the last blocks to be committed. If there is a correct replica that received $N-F$ votes for $r$, then it means that there are $N-F$ replicas sent their votes in the same round. Then, after network partitions or node failures, some replicas increase their rounds number $r+k$. However, the round $r$ is still the last round that have the parent blocks for the next round. For simplicity, we assume the leader $l_{i+1}$ is in $r$ and has proposed $B_{j+2}$. A leader, say $l_{i+2}$, received a $N-F$ votes for $B_{j+1}$. The latest valid votes in leader $l_{i+1}$ should be $B_{j+1}$ in $r$. This is because the leaders cannot accept any proposal without the proof of the previous block votes.

\subsection{Liveness}

BigBFT guarantees its liveness under the partial synchronous model. After GST, when network conditions are good, the network is stable and delay time is known. If there are less than $F\!\!<\!\! \frac{N}{3}$ Byzantine leaders, the honest leaders can reach a decision in a $\delta$ time. 

\textbf{Lemma 3} To ensure liveness, each round has a coordinator, a set of leaders, and the round number is incremented. The protocol after GST, with a correct coordinators and leaders, eventually become synchronized and the block will be added to the blockchain. 

$Proof.$ Let us assume that we have a correct coordinator called $C$, and leaders $L_{s}$ for round $r$ at time $t$. This means that $L_{s}$ have received the coordinator message $RoundQC$ that at least $N-F$ leaders signed in $r$ where $F<\frac{N}{3}$. Thus, at least $F+1$ correct leaders assign $C$ as a coordinator at time $t$. By time $t+\delta$, all correct leaders complete the protocol phases.

\textbf{Lemma 4} A new leaders set is chosen based on the their stakes. If a leader does not make progress, the leader $L_{i}$ will be discarded from the leader set $L_{s}$. 

$Proof.$ At the beginning of the round, the leaders is chosen by coordination phase. As per assumption all correct leaders are in the same round, therefore the correct leader will propose a prepare message with proof containing the latest votes $AggQC$. Since all leaders are in the same round, therefore leaders who are not successfully complete prepare and vote will eliminate from the $L_{s}$.

%% file: Evaluation.tex
\section{Performance Evaluation}
\label{sec:Evaluation}
\subsection{Implementation and setup}

\begin{figure}[!h]
	\centering
	\includegraphics[width=3.5in]{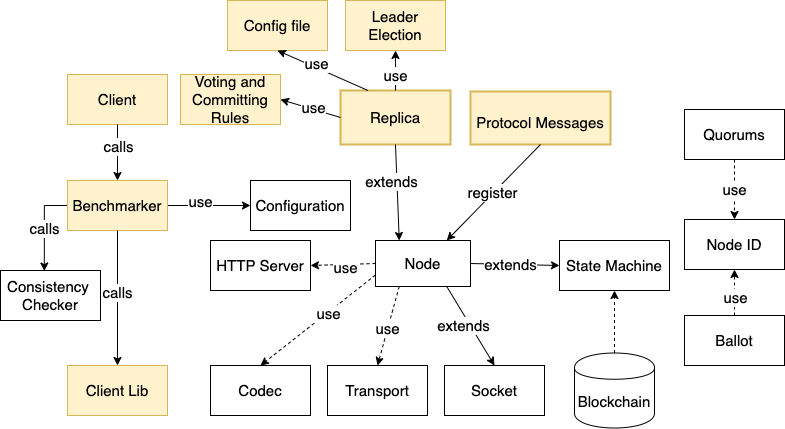}
	\caption{\textbf{The PaxiBFT architecture}}
	\label{fig:Paxibft}
\end{figure}

We implemented BigBFT in Go using PaxiBFT framework~\cite{PaxiBFT}, which uses core network files from Paxi~\cite{PAXI}. PaxiBFT is an open source for prototyping, evaluating, and benchmarking BFT consensus and replication protocols. As shown in Figure~\ref{fig:Paxibft}, PaxiBFT readily provides most functionality that any coordination protocol needs for replication protocols. The entire protocol is available as open source at \url{https://github.com/salemmohammed/BigBFT}.

We evaluate the performance of our BigBFT prototype protocol and compare it with PBFT~\cite{PBFT}, Tendermint~\cite{Tendermint}, Streamlet~\cite{Streamlet}, and HotStuff~\cite{Hotstuff}, using the Amazon EC2 instances. We chose both Wide Area Network(WAN) across 5 AWS regions(Ohio, N.California, Oregon, N.Virginia, and Canada) and Local Area Network(LAN).

We deployed BigBFT on up to 20 m5a.large virtual machines, each of which has 2 vCPU, 8GiB RAM, and 10Gbps network throughput. To ensure that the client performance does not impact the results, we used the larger m5a.xlarge instances with 4 vCPUs for the clients. Based on our experiments results, the network size is appropriate to state and conclude our findings. To push system throughput, we varied the number of clients up to 90 clients and used a small message sizes. In our experiments, message size did not dominate consensus protocols performance, but the complexity of consensus protocols dominates the performance.

\begin{figure}[!h]
	\centering
	\includegraphics[width=3.5in]{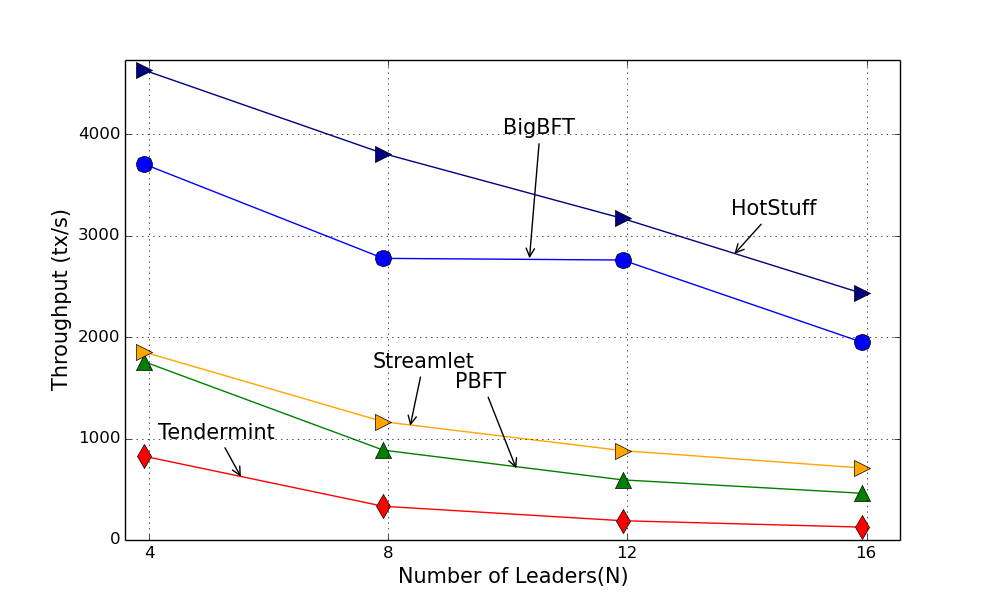}
	\caption{\textbf{Throughput comparison in LAN}}
	\label{fig:patterns}
\end{figure}

We compare the performance of protocols when $F=0$ and $N$ ranges from 4 to 20 full nodes. The results are shown in Figures~\ref{fig:patterns} and~\ref{fig:WANT}. In each graph, the Y-axis shows the throughput in tx/sec, and X-axis the number of nodes(N). We define the throughput as the number of transactions per second (tx/s for short) that leaders commit. As we can see in both Figures~\ref{fig:patterns} and~\ref{fig:WANT}, BigBFT achieves a better performance than PBFT, Tendermint, and Streamlet in LAN deployment. In a comparison to HotStuff, BigBFT is close to Hotstuff's throughput but has better latency. This is because, in both wide and local environment, the network is the bottleneck and the message patterns of BFT protocols, namely PBFT, Tendermint, and Streamlet, tend to be expensive. BigBFT on the other hand maintains a low number of message latency for each request to be committed and simple message patterns. 

The latency messages are 4 messages. However, Hotstuff has 10 messages regardless of chain protocol as we summarized in Table~\ref{tab:template}. These latency messages are not important in LAN due to the short distance between nodes. As a result, BigBFT performs well in LAN deployment as you can see in Figure~\ref{fig:patterns}. In contrast to LAN, there is a long delay in WAN between nodes that helps BigBFT due to its low number of message latency to commit the requests faster. 
Compared with PBFT, Tendermint, Streamlet, and Hotstuff, BigBFT puts less stress on the leaders. Moreover, the performance of BigBFT is very close to that of HotStuff and Paxos~\cite{alqahtani2021bottlenecks}. 

Hotstuff has point to point communication while BigBFT is still using broadcasting topology between nodes even though we amortized the messages over the number of requests. The combination of messages in the vote phase causes some delay, but that delay was unnoticed in WAN due to the distance/latency gain from sending messages in parallel.

\begin{figure}[!h]
	\centering
	\includegraphics[width=3.5in]{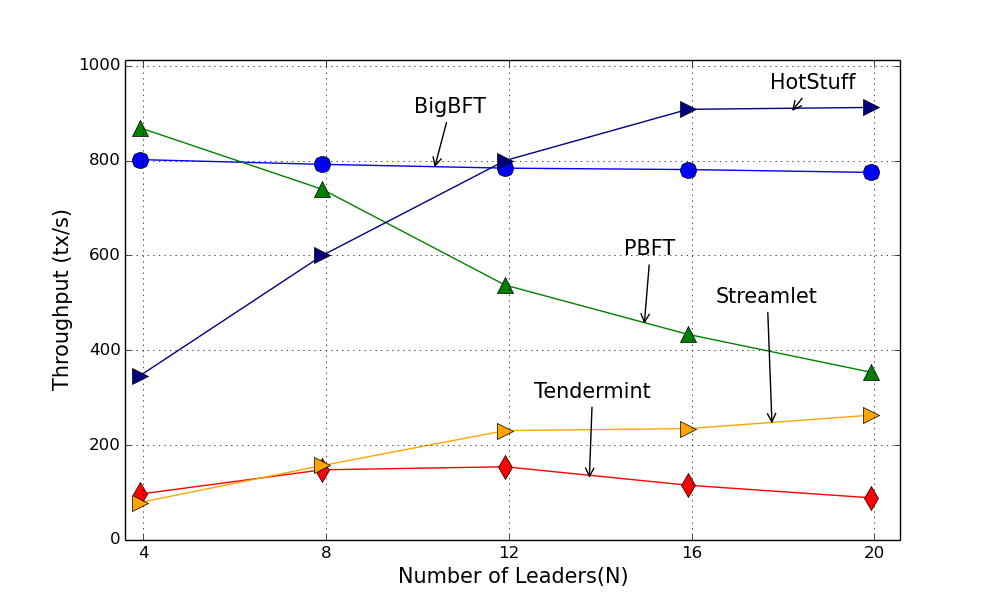}
	\caption{\textbf{WAN's throughput comparison in Virginia, California, Oregon, Ohio, and Canada}}
	\label{fig:WANT}
\end{figure}

\begin{figure}[!h]
	\centering
	\includegraphics[width=3.5in]{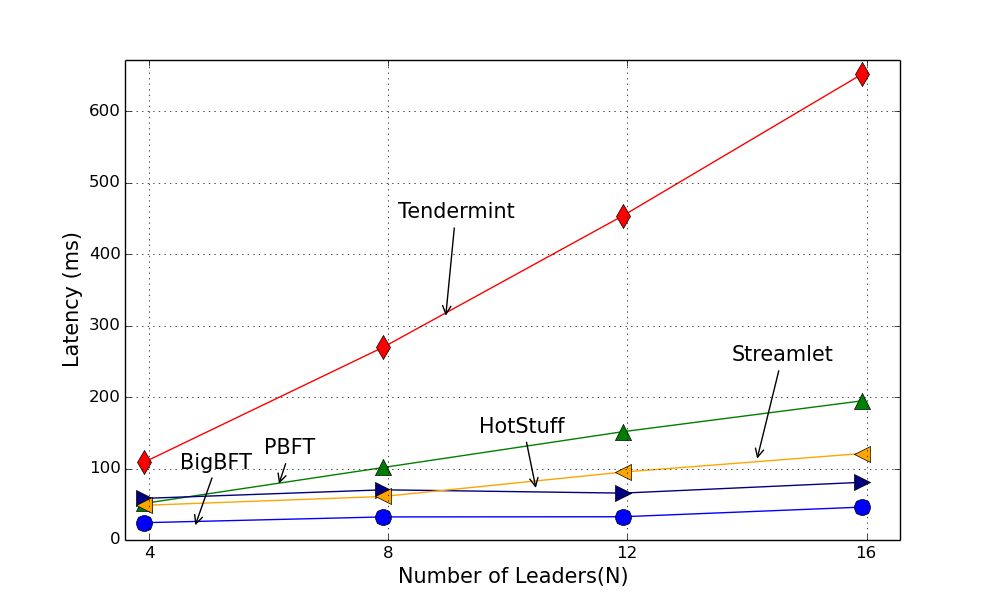}
	\caption{\textbf{Latency comparison in LAN}}
	\label{fig:WANT}
\end{figure}

\subsection{Performance comparison between single and multi-leaders protocols}  
\label{sec:COM}

To study the performance of BigBFT, we compare BigBFT latency and throughput with PBFT~\cite{PBFT}, Tendermint~\cite{Tendermint}, Hotstuff~\cite{Hotstuff}, and Streamlet~\cite{Streamlet}. We run all of these protocols with no faults to make sure we capture their absolute best performances. We chose these protocols to compare with BigBFT because we had studied and analyzed them in our previous work~\cite{alqahtani2021bottlenecks}. As a result of our previous work, we introduced BigBFT to alleviate the bottlenecks. BigBFT relies on a two phase common case commit protocol with $3F+1$ replicas illustrated in Figure~\ref{fig:BigBFT}. To provide a fair comparison, all protocols implemented on the same framework PaxiBFT~\ref{fig:Paxibft}. 

\begin{figure}[!h]
	\centering
	\includegraphics[width=3.5in]{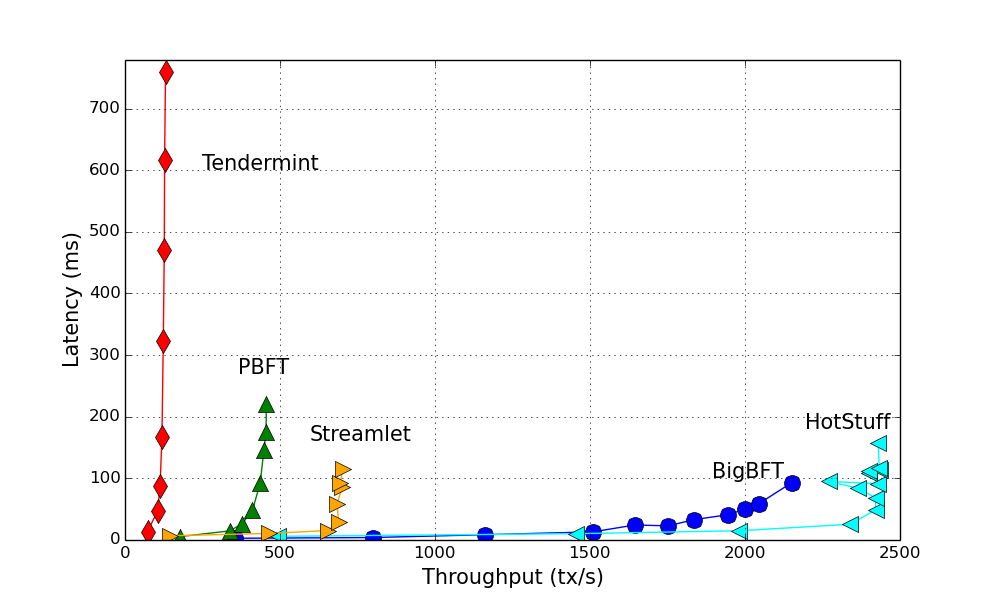}
	\caption{\textbf{System throughput and the latency on 20-node LAN cluster}}
	\label{fig:LANC}
\end{figure}

\begin{figure}[!h]
	\centering
	\includegraphics[width=3.5in]{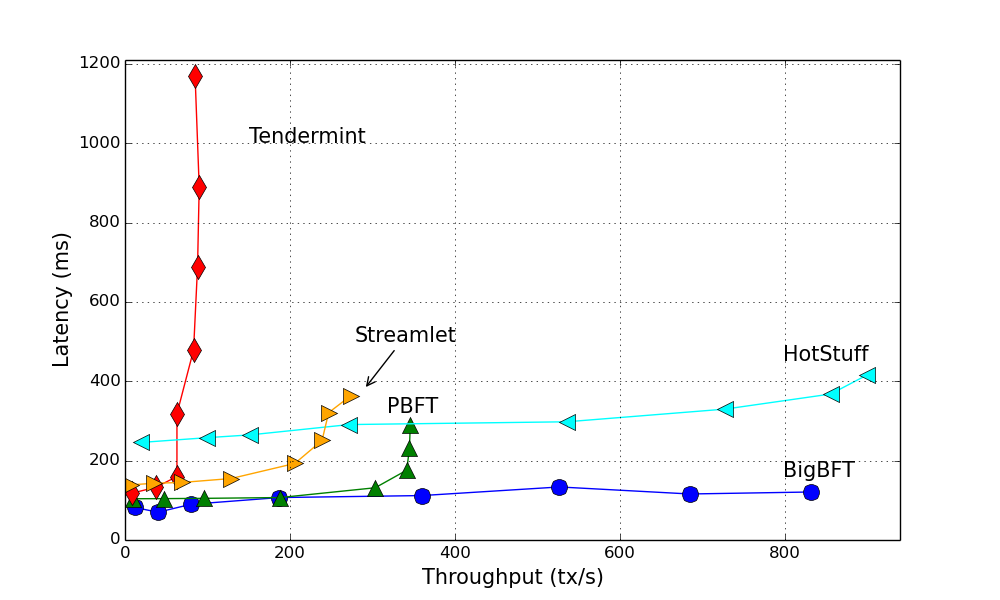}
	\caption{\textbf{System throughput and the latency on 20-node WAN cluster in Virginia, California, Oregon, Ohio, and Canada}}
	\label{fig:WANC}
\end{figure}

Figures~\ref{fig:LANC} and~\ref{fig:WANC} illustrate the latency versus throughput performance of these five protocols in a 20-node cluster. In each graph, the X-axis shows the throughput in tx/sec, and Y-axis the latency in ms. BigBFT and HotStuff did not saturated very quickly in Figure~\ref{fig:WANC}. Pushing the system throughput to its limit is difficult in WAN. In LAN, pushing the system throughput to its limit to get the system bottlenecks is easy due to the short network pipe between instances.

At this cluster size, BigBFT shows better latency than other protocols because it allows parallel executions. This removes the single leader bottleneck and allows BigBFT to have more throughput than many protocols. On the other hand, PBFT, Tendermint, and Streamlet are significantly limited by their increased communication costs.

PBFT is limited by a single leader and quadratic exchanging messages for every request. PBFT gets saturated quickly and reaches its limit of around 480 requests per second in LAN and 360 requests per second in WAN. While BigBFT have higher latency in LAN, but we see that it scales to higher throughput than other protocols except Hotstuff with better latency than all protocols. BigBFT provides great throughput improvements over traditional PBFT in wide area deployments. 

BigBFT maintains low latency for much higher levels of throughput and shows higher throughput than Tendermint and Streamlet in WAN. Tendermint gets saturated quicker by a more complicated messaging and processing despite our workload having no conflicts.

\subsection{System Payload Size}  
\label{sec:COM}

\begin{figure}[!h]
	\centering
	\includegraphics[width=3.5in]{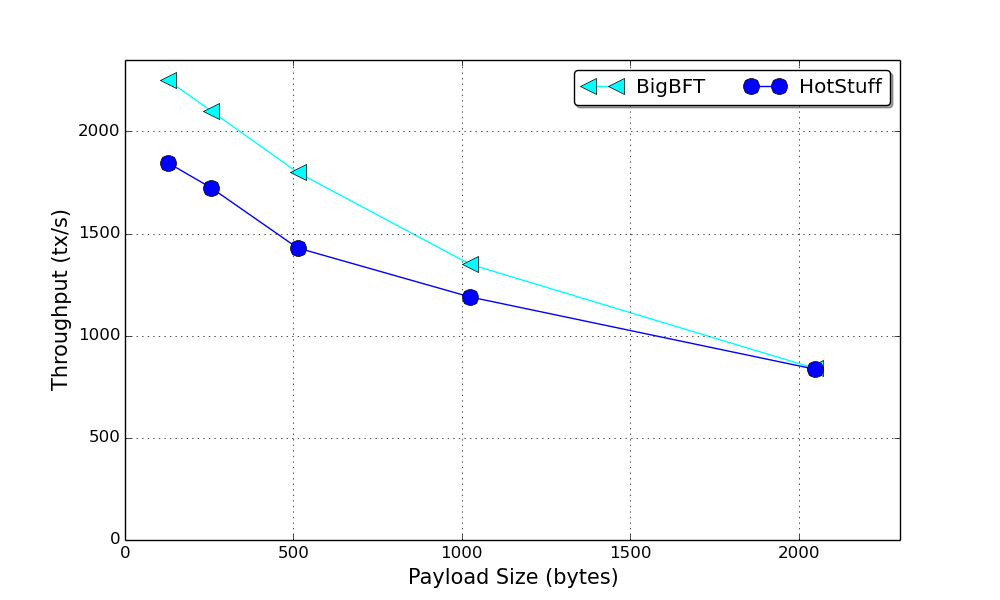}
	\caption{\textbf{Maximum throughput at various payload sizes}}
	\label{fig:payload}
\end{figure}

Payload size have an impact on the communication performance of the system. With the large messages, system requires more resources for serialization and more network capacity for transmission. To study how different payload size impacts the performance of our studied protocols and BigBFT, we experiment with 16 node clusters.  We measured the maximum throughput on each system under a write-only workload by 90 clients. Figure~\ref{fig:payload} shows the maximum throughput of BigBFT and Hotstuff at payload sizes varying from 128 to 2048 bytes. While BigBFT show less throughput than HotStuff at the beginning of payload sizes, both protocols exhibit a similar relative level of degradation as the payload size increases.

%% file: Discussion.tex
\section{Analysis and Discussion}
\label{sec:anal}

\begin{table*}
\vspace*{-4mm}
\centering
\begin{tabular}{ |l|c|c|c|c|c|c|c| } 
 \hline
&PBFT~\cite{PBFT}&HotStuff~\cite{Hotstuff}&Mir-BFT~\cite{Mir}&Streamlet~\cite{Streamlet}&Tendermint~\cite{Tendermint}&BigBFT(This paper)\\
\hline
Critical path & 5 & 10& 5 & 4 & 5& 4\\
\hline
Normal Message Complexity & $O(N^2)$ & $O(N)$ & $O(N^2)$   & $O(N^3)$  &  $O(N^2)$ & $O(N)$\\
\hline
Multiple View Change & $O(N^4)$ & $O(N^2)$ & $O(N^4)$ & $O(N^4)$& $O(N^3)$ & $O(N)$\\
\hline
Responsive & Yes & Yes & Yes & No & No & Yes \\
\hline
\end{tabular}
\caption{\textbf{Characteristics of BFT consensus protocols}}
\label{tab:template}
\vspace*{-3mm}
\end{table*}

In this section, we compare the strengths and weaknesses of the BigBFT and provide back-of-the-envelope calculations for estimating the latency and throughput performance. Table~\ref{tab:template} provides a synopsis of the blockchain protocols characteristics we compared with BigBFT. We elaborate on these next.

\textbf{Time Complexity.} In Mir-BFT, and PBFT, the normal executions have a quadratic complexity. When the leader is a malicious, the protocol changes the view with a different leader using a view-change which contains at least $2F+1$ signed messages. Then, a new leader broadcasts a new-view message including the proof of $2F+1$ signed view-change messages. Leaders will check the new-view message and broadcast it to have a match of $2F+1$ new-view message. The view-change has then $O(N^3)$ complexity and $O(N^4)$ in a cascading failure. 

Tendermint reduces the message complexity that is caused by view-change in PBFT, to a total $O(N^3)$ messages in the worst case. Since at each epoch all leaders broadcast messages, it happens that during one epoch the protocol uses $O(N^2)$ messages. Thus, in the worst case scenario when there is $F$ faulty leaders, the message complexity is $O(N^3)$~\cite{Tendermint}.

Streamlet has communication message complexity $O(N^3)$. Streamlet loses linear communication complexity due to all-to-all communication in vote message. In the worst case scenario when there is a leader cascading failure, the Streamlet message complexity is $O(N^4)$. 

HotStuff all have linear message complexity. The worse case communication cost in these protocols is $O(N^2)$ considering worst-case consecutive view-changes. 

BigBFT has communication message complexity $O(N)$. BigBFT has a linear communication complexity in vote phase in the best case scenarios. In the worst case scenario when there is a leader cascading failure, the BigBFT message complexity does not change because overlapping between propose-vote phases and coordination phase.
 
\subsection{Load and Capacity}

BigBFT protocol reaches consensus once a quorum of participants agrees on the same decision. A quorum can be defined as sets containing $N-F$ majority leaders in the system with every pairs of set has a non-empty intersection. To select quorums $Q$, quorum system has a strategy $S$ in place to do that. The strategy leads to a load on each validator. The load $\ell(S)$ is the minimum load on the busiest leader. The capacity $Cap(S)$ is the highest number of quorum accesses that the system can possibly handle $Cap(S)=\frac{1}{\ell(S)}$. 

In single leader protocols, the busiest node is the leader. In BigBFT, all nodes are busy all the time.

\begin{equation}\label{eq1}
\begin{split}
\ell(S) & = \frac{1}{L}(Q-1)NumQ + (1-\frac{1}{L})(Q-1) \\
 \end{split}
\end{equation}

where $Q$ is the quorum size chosen in both leader and followers, NumQ is quorums number handled by leader/follower for every block request, and $L$ is the number of operation leaders. There is a $\frac{1}{L}$ chance the node is the leader of a request. Leader communicates with $N-1=Q$ nodes. The probability of the node being a follower is $1- \frac{1}{L}$, where it only handles one received message in the best case. In the equations below, we present the simplified form of BigBFT and PBFT protocols, and calculate the result for $N = 9$ leaders. The protocols perform better as the load decreases.

\begin{equation}\label{eq2}
\begin{split}
\ell(BigBFT) & = 5 \frac{5}{9}
\end{split}
\end{equation}

In BigBFT protocol, equation~\ref{eq2} with $N$ leaders, and L = N, quorum size Q = $\lfloor{\frac{2N}{3}}\rfloor$, and number of quorums $NumQ = 2$.
Many requests $R$ have overlapped in $Q$ and $NumQ$ due to parallel executions. Therefore, we divide $Q$ and $NumQ$ over number of requests $R$ which equals to number of active leaders $N=R$. 
PBFT is a single leader protocol with $Q = \lfloor{\frac{2N}{3}}\rfloor$ and $NumQ = 2$. The load on PBFT is $\ell(PBFT) = 10$

\subsection{Latency}

The formula~\ref{eq5} calculates the latency of BigBFT.
\begin{equation}
\begin{split}\label{eq5}
Latency(S) = Critical\; Path  + D_{L} + \delta
\end{split}
\end{equation}

The critical path denotes the number of one-way message delays. BigBFT's critical path has a 4-message delay as illustrated in Table~\ref{tab:template} for multiple consensus. $D_{L}$ is the round trip message between a client and designated leader. In Table~\ref{tab:template}, PBFT has a 5-message delay for single consensus.

%% file: Futurework.tex
\section{Future Work}
\label{sec:fw}

\subsection{Improving Scalability by Enabling Local Communication}

One direction for future work is to enable hierarchical communications in order to scale the protocol to more nodes. At each round, a global coordinator selects a local coordinator in each region to maintain the configuration blockchain and to restrict the communication within each region. The global coordinator will reach consensus with local coordinators on sharding the space between regions. Local coordinator will further shard the space among all local leaders. Then, each region will run BigBFT locally among nodes to order blocks.

\subsection{Client as a Coordinator}

We are planning to enable client node to act as a system coordinator to further improve BFT scalability by reducing the interactions between replicas. The client can assign the request's sequence number, choose a set of trusted replicas to form a quorum, and learn the status of the request from the chain if committed. For instance, the client may choose a quorum of nodes including well-known validators, as shown in Stellar protocol~\cite{mazieres2015stellar}, with the high stake values to commit its blocks. Client may use blockchain to check recent honest validators that are able to commit and add to the chain. For more flexibility, the client might develop some set of policies to choose its quorum such as choosing common nodes between recent last two committed blocks. However, the system should have a security mechanism to prevent a malicious client from violating the safety of the system.

%% file: Conclusion.tex
\section{Concluding Remark}
\label{sec:concl}

We have presented BigBFT, a BFT-based consensus protocol for blockchains.
BigBFT alleviates the communication bottlenecks in single leader BFT protocols by enabling multi-leader executions, and reduces the number of communication phases to be only two for reaching consensus on proposed blocks. This is achieved by pipelining blocks both within a round and across rounds. BigBFT also decouples the coordination phase from the main protocol and pipelines coordination round with consensus rounds to improve the system performance. We analyzed BigBFT performance by using a load formula and compared it with PBFT, Streamlet, Tendermint, and Hotstuff protocols. Our experimental evaluations show BigBFT's advantages in latency or throughput over other protocols. 




%% file: Akc.tex
\section{Acknowledgements}
This project is in part sponsored by the National Science Foundation (NSF) under award number CNS-2008243.